\documentclass[useAMS,usenatbib]{mn2e}
\usepackage{latexsym,graphicx,natbib}

\newcommand{\be} {\begin{equation}}

\newcommand{\ee} {\end{equation}}

\newcommand{\bc}{\begin{center}}
\newcommand{\ec}{\end{center}}
\def\ltsima{$\; \buildrel < \over \sim \;$}
\def\simless{\lower.5ex\hbox{\ltsima}} 
\def\loe{\lower.5ex\hbox{\ltsima}}
\def\gtsima{$\; \buildrel > \over \sim \;$}
\def\simgreat{\lower.5ex\hbox{\gtsima}}
\def\goe{\lower.5ex\hbox{\gtsima}}

\def\ergscm2 {erg\,s$^{-1}$cm$^{-2}$}

\def\msun {M_{\odot}}
\def\ergs{{\rm\,erg\,s^{-1}}}

\title[The BH in NGC1313 X-2]{The black hole in NGC 1313 X-2: constraints on the mass from optical observations}

\author[Patruno \& Zampieri]
{
Alessandro Patruno$^1$ \& Luca Zampieri$^2$ \\
$^1$Astronomical Institute `A. Pannekoek', Univeristy of Amsterdam,
Science Park 904, 1098 XH Amsterdam, The Netherlands: \tt{a.patruno@uva.nl}\\
$^2$INAF - Osservatorio Astronomico di Padova, Vicolo
dell'osservatorio 5, 35122 Padova, Italy :{ \tt luca.zampieri@oapd.inaf.it}}

\begin{document}

\pubyear{2009}

\maketitle

\label{firstpage}

\begin{abstract}

We present a theoretical study on the nature of the ultra-luminous
X-ray source NGC 1313 X-2. We evolved a set of binaries with high mass
donor stars orbiting a $20 M_\odot$ or a $50-100 M_\odot$ black
hole. Using constraints from optical observations we restricted the
candidate binary system for NGC 1313 X-2 to be either a $50-100
M_\odot$ black hole accreting from a 12--15$\msun$ main sequence star or a
$\sim20 M_\odot$ black hole with a 12--15$\msun$ giant donor. If the
modulation of $6.12\pm0.16$ days recently identified as the orbital
period of the system is confirmed, a $\sim20\msun$ black hole model
becomes unlikely and we are left with the only possibility that the
compact accretor in NGC 1313 X-2 is a massive black hole of $\sim
50-100\msun$.

\end{abstract}


\section{Introduction}\label{Intro}

Ultra luminous X-ray sources (ULXs) are off-nuclear extragalactic
X-ray sources with isotropic bolometric luminosities in excess of the
Eddington limit for a $\sim20\msun$ black hole (henceforth referred to
as BH). An empirical approach defines ULXs as sources with X-ray
luminosities in the range $L_{X}\sim10^{39}-10^{41}\ergs$.  These
sources are quite common in starburst and late type galaxies and are
thought to be mostly binaries with a donor star transferring matter
onto a black hole. There is not yet a general consensus on the
mechanism that drives such large luminosities (see
e.g. \citealt{zam09} and references therein). Some proposed
explanations involve $\simless 20\msun$ BHs with thick discs producing
beamed emission, slim discs with photon bubble trapping and
super-Eddington emission (see e.g., \citealt{pou07} and references
therein), or two-phase super-Eddington radiatively efficient discs
\citep{soc06}. The most intriguing possibility would be the presence
of the so-called intermediate mass black holes (IMBHs): even if ULXs
are genuine isotropic emitters, the Eddington limit would not be
violated if the accretor has a mass of
$\sim10^{2}-10^{4}\,M_{\odot}$. An alternative formation scenario has
been proposed and recently explored in detail in which a
portion of ULXs contains $\sim 30-90 M_\odot$ BHs formed in a low
metallicity environment and accreting in a slightly critical regime
(\citealt{map09,zam09}). The binary nature of ULXs was supported by
the possible orbital periodicity at 62 days detected by \citet{kaa06,
kaa06b} and \citet{kaa07} for the ULX M82 X-1. Constraints from the
orbital period, X-ray luminosity and optical photometry suggested that
the most likely explanation for the compact accretor in this ULX is an
IMBH of mass larger than 200 $M_{\odot}$ (\citealt{pat06}; see however
\citealt{beg06} for an alternative interpretation without an IMBH).
\citet{pat08} and \citet{mad08} demonstrated how several further
constraints can be put on the nature of ULXs from the identification
of optical counterparts. In the majority of cases optical
identifications indicate the existence of high mass donor stars
($M\simgreat 8 M_{\odot}$). The optical emission coming from the donor
is expected to be strongly contaminated by the contribution of the
external regions of the accretion disc, and by reprocessing of the
X-ray radiation generated in the innermost portion of the disc.
Contribution from the X-ray irradiation of the donor star surface also
plays a role (\citealt{cop05}, \citealt{pat08}).

The identification of a unique optical counterpart however is not an
easy task, as ULXs are often observed in crowded regions of star
formation.  The ULX NGC1313 X-2 (henceforth referred to as X-2) is one
of the most promising sources in this respect. \citet{zam04} and
\citet{muc05} first identified two candidate counterparts for this
source (C1 and C2).  \citet{muc07} and \citet{liu07} pinpointed C1 as
the most likely counterpart by means of a model of the optical
emission and a refined analysis of the HST and {\it Chandra}
astrometry. Through an independent theoretical investigation, we
showed that the object C1 is the only one consistent with the
properties predicted by a binary evolution model
(\citealt{pat08}). \citet{muc05} estimated a mass of $\sim
20\,M_{\odot}$ for C1, later refined to be in the interval 10-18
$M_\odot$ by \citet{muc07}. \citet{gri08} identified the donor as a
star of 8-16$\msun$ whereas \citet{liu07} proposed a somewhat smaller
mass ($\sim 8\msun$). \citet{gri08} derived an age of $20\pm5$ Myr for
the star cluster where X-2 resides and hence an upper limit for the
donor mass of $\sim 12\msun$.  However, all these observational
studies did not consider the effects of binary evolution on the donor
star colours and age. Taking into account the effects of binary
evolution and irradiation, we estimated the mass of C1 to be $\simless
15\,M_{\odot}$ (\citealt{pat08}). Finally, from an estimate of the
amount of energy injected in the bubble nebula surrounding this ULX,
\citet{pak06} derived a timescale for the active phase of X-2 of the
order of $\sim 10^{6}$ yr.  \citet{pak06} further confirmed C1 as the
likely counterpart reporting evidence of a broad 4686 He II line in
the optical spectrum of X-2.

Recently, \citet{liu09} tentatively identified a modulation in the $B$
band lightcurve of X-2 with a period of $6.12\pm0.16$ d. If this is
confirmed, X-2 will be the most constrained ULX known to date.
In this Letter we will use all the available data of X-2 coming from
optical observations (including the $\sim 6$ d orbital period) and
compare them with the evolution of an ensemble of irradiated X-ray
binary models in order to constrain the nature of the compact
accretor.

\begin{figure*}
\begin{center}
\rotatebox{-90}{\includegraphics[width=5.5cm]{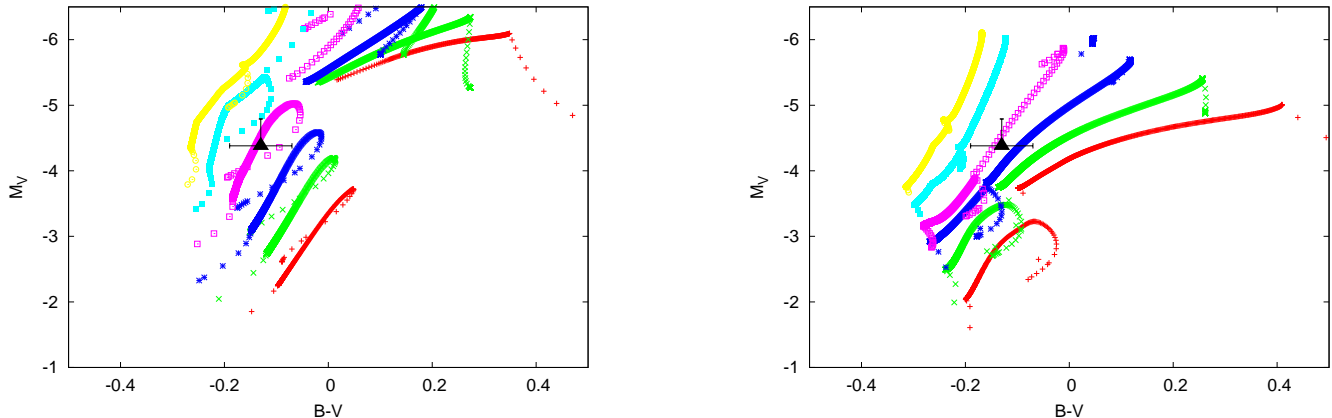}}
\end{center}
\caption{Colour-Magnitude Diagram (CMD) for binaries with a 100$\msun$
({\it left panel}) and a 20$\msun$ BH ({\it right panel}), undergoing
\textbf{case AB} mass transfer. The black triangle corresponds to the
optical counterpart C1 with its 1$\sigma$
errorbar. The assumed distance for X-2 is 3.7 Mpc and the errorbar in
$M_{V}$ reflects the maximum uncertainty in the distance determination of
NGC 1313 (see text). All the tracks are plotted only during the
contact phases. The effects of irradiation and the optical
contamination of the accretion disc are included. The
donors are: $25 \msun$ ({\it yellow}), $20 \msun$ ({\it cyan}), $15
\msun$ ({\it pink}), $12 \msun$ ({\it blue}), $10 \msun$ ({\it
green}), and $8 \msun$ ({\it red}).}
\label{caseAB}
\end{figure*}

\section{Binary evolution and X-ray reprocessing}\label{evol}

The binary evolution model adopted here is the same as outlined in
\citealt{pat08} (to which we refer for details; see also
\citealt{mad08}, \citealt{pat06}, \citealt{mad06}, \citealt{rap05}, \citealt{pat05}, \citealt{pod03}, for a discussion on BH
binary evolution models).  We consider two ensembles of binaries: the
first with BHs of $20\,M_{\odot}$ and the second with BHs of
$100\,M_{\odot}$.  The donor stars have initial masses between 8 and
25$M_{\odot}$. Two different values of the (zero age) metallicity,
both sub-solar, are considered: $Z=0.004$ (\citealt{ryd93}) and
$Z=0.01$ (close to the value $Z=0.008$ estimated by \citealt{had07}
and \citealt{wal97}).  If the donor starts a contact phase via
Roche-lobe overflow (RLOF), we assume that a geometrically thin
optically thick accretion disc forms around the BH. The efficiency for
the conversion of rest mass into radiation is fixed at $10\%$. The
accretion rate is instantaneously taken to be equal to the
mass-transfer rate from the companion. When the accretion rate ${\dot
M}$ exceeds the Eddington rate ${\dot M}_{Edd}$, we impose ${\dot M} =
{\dot M}_{Edd}$ and assume that the excess mass is expelled from the
system.  In our simulations, all the binaries that start RLOF on the
main sequence (MS) have also a second episode of mass transfer after
the terminal age main sequence (TAMS). Therefore we term such binary
models as \textbf{case AB} while those starting the first contact
phase after the TAMS are termed \textbf{case B}.

The UV/optical luminosity is computed summing to the donor emission
the contribution of the accretion disc, and including the effects of
the reprocessed X-ray radiation (see \citealt{pat08} for details). In
the present version of the code we improved the accuracy in the
calculations of the optical colours of the irradiated accretion disc,
which are now $\sim 0.1$ mag redder than those previously reported.
None of the results and conclusions published in \cite{pat08}
are significantly affected by this revision. 

For all the donors considered here the disc contribution to the
UV/optical emission can be dominant for BHs of $\sim 100 M_\odot$,
whereas it is less important but still significant for 20$\msun$ BHs
\citep{pat08}. The albedo of the irradiated layers is fixed at
$f_a=0.9$. To investigate the effects of a different albedo we evolved
some binaries with $f_a=0.7$ and $f_a=0.95$. As reference, we chose
five values for the inclination angle of the binary, $i=0^0, 30^0,
45^0, 60^0, 80^0$. For each angle we computed the optical emission
summing the average luminosity of the irradiated and non-irradiated
surfaces of the donor to the flux emitted from the visible part of the
accretion disc.

\section{Observational constraints}\label{obs}

X-2 is located in the barred spiral galaxy NGC 1313 at a distance of
3.7-4.27 Mpc (\citealt{tul88}, \citealt{men02}, \citealt{riz07}). Its
observed X-ray luminosity varies between a few $\times 10^{39}\ergs$
and $3\times10^{40}\ergs$ in the 0.3-10 keV band
(\citealt{fen06,muc07}). 
Recently, \citet{liu09} found a possible periodicity of $6.12\pm0.16$
d in the $B$ band lightcurve of C1, that was interpreted as the
orbital period of the binary. Three cycles were detected in the $B$
band, while no modulation was found in $V$.  According to
\citet{liu09}, the period is $12.24\pm0.16$ d if the X-ray irradiation
of the donor is unimportant, while it is $6.12\pm0.16$ d in case
of significant irradiation. Previous studies carried
out on the available \textit{HST} and \textit{VLT} observations led to
negative results \citep{gri08}. More recently, lack of significant
photometric variability on a new sequence of \textit{VLT} observations
has been reported by \cite{gri09}. Therefore, we consider the
detection of \citet{liu09} with caution and are aware that it will
need to be confirmed before any definite conclusion can be drawn.

In the following we will use the $V$ and $B$ band photometry of C1 as
determined by \citet{muc07}.  As the source is variable, we have
further corrected the magnitudes and colours by using the average value of $V$
and by propagating the errors on $V$ and $B$. The error in
the absolute magnitudes is taken to be equal to the maximum
uncertainty in the different distance determinations of NGC 1313
(\citealt{tul88,men02,riz07}). Concerning
the reddening, two different estimates of the colour excess were
derived in the literature: $E(B-V)=0.1$ \citep{muc07,gri08} and
$E(B-V)=0.3$ \citep{liu07}. As the analyses of \cite{muc07} and
\cite{gri08}, based in part on independent methods 
converge toward the same value, in the following we chose
$E(B-V)=0.1$\footnote{If we consider E(B-V)=0.3, all the models with a
20$\msun$ BH become incompatible with the position of C1, and the
minimum donor mass required to match the observations for a 100$\msun$
BH becomes $M\simgreat 25\msun$ with characteristic age $t\simless
10$Myr (H-shell burning phase), in strong disagreement with the
observed cluster age of $20\pm 5$Myr, reported by \citet{gri09}}. The
adopted $M_{V}$ magnitude is therefore in the range -4.38 to -4.79 and
the $B-V$ colour is $-0.13\pm0.06$.\\

\section{Results}\label{results}

\subsection{Case AB mass transfer}
In Fig.~\ref{caseAB} we show the results of the binary evolution calculations
(including the effects of irradiation) for Z=0.01 and \textbf{case AB} mass 
transfer, and for an inclination $i=0^0$. 
The track for a 15$\msun$ donor with a $100 \msun$ BH passes through
the C1 errorbox during the MS, whereas the 20$\msun$ BH matches the
observations during the giant phase. When this happens, the orbital
period for the 15$\msun$ donor and $100 \msun$ BH model is $\sim 5.9$
d, very close to the determination of \citet{liu09}. The donor age is
$\sim 16$ Myr, which is consistent with the estimate of \citet{gri08}
for the host OB association.

We decreased the BH mass to 50--70$\msun$, keeping the donor mass
at 15$\msun$, to verify if a slightly lighter BH would produce a
significantly different result. In all cases the donor colours, age
and orbital period match the observations when the star is close to
the TAMS, with values similar to those of
the 100 $\msun$ BH case.
The mass transfer of the 15$\msun$ donor at the position of C1 is
typically between $\dot{M}\sim2\times 10^{-7}\msun\rm\,yr^{-1}$ and
$\dot{M}\sim2\times 10^{-6}\msun\rm\,yr^{-1}$. This is consistent with
the measured average luminosity of X-2, $\sim 4 \times
10^{39}\rm\,erg\,s^{-1}$ \citep{muc07}, which requires $\dot{M}\sim
10^{-6}\msun\rm\,yr^{-1}$.
When considering the tracks with a smaller metallicity (Z=0.004), the
situation is similar and the effect on the stellar colours is minimal. 
Therefore we will consider for reference only tracks calculated for Z=0.01.

The $M_{V}$ band magnitude of ULXs is lower (i.e., the luminosity larger)
for 50-100$\msun$ BHs than for 20$\msun$ BHs (see Fig.~\ref{caseAB} and
\citealt{pat08}). This is mainly due to the larger optical
contamination of the irradiated disc and for an intrinsic binary
evolution effect that makes a companion star around a 50--100$\msun$ BH
brighter than that around a 20$\msun$ BH when RLOF starts at a
fixed donor age. For a 50--100$\msun$ BH the (irradiated) disc flux exceeds
the (irradiated) donor flux up to 100-300\%, depending on the
photometric band and the mass ratio considered, while for a 20$\msun$
BH it is between 20 and 130\% of the donor flux at maximum
accretion rate.  A small increase of the order of $\sim 0.1$ mag in
$B-V$ occurs when increasing the BH mass.

All the results remain essentially unchanged if we increase the
inclination angle up to $80^0$. Only for $i\simgreat 80^0$ there is a
significant decrease of the accretion disc contribution and hence of
the luminosity ($\propto \cos i$), which is particularly significant
for binaries around a 50--100$\msun$ BH.  A light curve folded over
the orbital period was simulated calculating the area of the
irradiated/non-irradiated surface seen by a distant observer as a
function of orbital phase and inclination, integrating the specific
intensity separately over the two surfaces and summing up the
resulting fluxes. At large inclination angles ($i\simgreat 60^0$) the
light curve is approximately sinusoidal and the maximum amplitude of
the orbital modulation in the $B$ band is $\sim 0.04$ mag (including
disc emission), about a factor of $\sim 2$ smaller than the amplitude
of the observed modulation \citep{liu09}. For a 20$\msun$ BH, the
contribution from the disc is less important and the amplitude of the
modulation is close to the observed value ($\sim 0.1$ mag). However,
if moderate beaming is assumed, the X-ray radiation is no longer able
to hit the donor surface and the orbital modulation caused by X-ray
heating is expected to be largely suppressed. To verify this scenario,
we evolved a set of tracks with a 20$\msun$ BH and no effect of
irradiation for \textbf{case AB}.
The tracks of a 15$\msun$ donor are compatible with C1 when
the star is on the giant branch with an age of $\sim19$ Myr, although the
period of the binary is $\sim 15-23$ d.

When considering binaries with irradiation, the evolutionary tracks do
not show dramatic changes in position and shape on the CMD when
varying the albedo $f_a$. The main effect is an increase (up to a
factor 2 for $f_a=0.7$) of the amplitude of the orbital modulation
caused by X-ray irradiation. The difference in the $B-V$ colour is
$\sim 0.1$ mag or less, while the change in the $M_{V}$ magnitude is
smaller than 1 mag.  When comparing the tracks with the position of C1
on the CMD, this translates into an uncertainty of only a few solar
masses in the determination of the donor mass. This is true also for
the {\bf case B} discussed below.  Both 20$\msun$ and 50--100$\msun$
BH models are still crossing C1 when the donor is on the giant branch
and on the MS, respectively.

By considering all these small sources of uncertainties for the
stellar tracks, we can consider the position of C1 also consistent
with donors on the MS with masses of 12--20$\msun$ and a
50--100$\msun$ BH.  However, when the track of the 20$\msun$ donor
crosses C1, the donor age is 4--11 Myr with a period 1.5--5.9 d.  The
donor is therefore too young to be compatible with the stellar cluster
age (15--25 Myr). For the 12$\msun$ donor, the age is 20--22 Myr and
the orbital period is 3.1--5.5 d.

As mentioned above, the photometry of C1 is not consistent with the
tracks relative to a $20 \msun$ BH for donors on the MS. For these
models, the mass transfer rate never exceeds the Eddington limit by
more than a factor 2 during the main sequence for donor masses
$M\simless 15\msun$. This means that even in case of genuinely
super-Eddington accretion, the amount of X-ray irradiation would never 
greatly exceed the value used in our calculations.

The same is true for the orbital period. At the TAMS, it increases
with the BH mass, and it is around 4.5--5 d for a BH of 20$\msun$ and
5--6 d for a 100$\msun$ BH. Therefore all the $20\msun$ BH binaries
need to be in the H-shell burning phase to be consistent with the
observations. The donors compatible with the position of C1 have ages
in the range 12--36 Myr and orbital periods of 5--20 d, with a mass
transfer rate always above the Eddington limit.  Donors of 10$\msun$
and 20$\msun$ can immediately be excluded since they are too old
($\sim 36$ Myr) or too young ($\simless12$ Myr) to be compatible with
the star cluster age (15--25 Myr).

\begin{figure*}
\begin{center}
\rotatebox{-90}{\includegraphics[width=5.5cm]{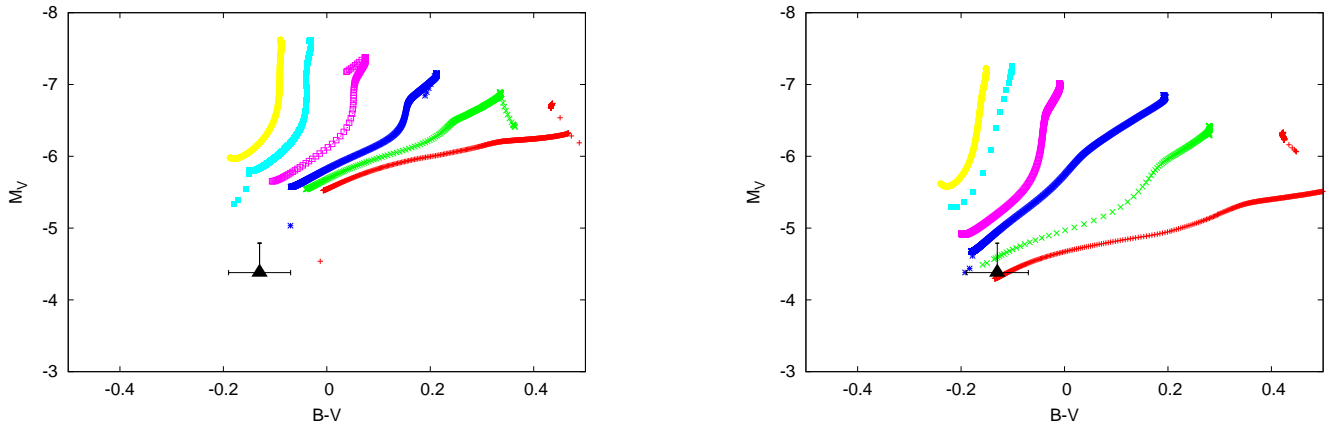}}
\end{center}
\caption{CMD for binaries with a BH of $100 \msun$ ({\it left panel}) and $20 \msun$ 
({\it right panel}) for \textbf{case B} mass transfer. The notation is the same as in 
Fig.~\ref{caseAB}. 
For a $100 \msun$ BH system the $M_{V}$ band magnitude of the tracks is too large for any model, 
while a 20$\msun$  BH with a donor of $\sim 10-12\msun$ is consistent with the 
position of the counterpart C1. However, these short-lived RLOF systems are not
consistent with the dynamical age inferred from the bubble nebula surrounding X-2.}
\label{caseB}
\end{figure*}

\subsection{Case B mass transfer}

In Fig.~\ref{caseB} we show the results of the binary evolution
calculations for \textbf{case B} mass transfer. We assume again
Z=0.01 and an inclination $i=0^0$. The RLOF starts when the donor
leaves the MS and its envelope expands crossing the
Hertzsprung-Gap. The figure shows that 50--100$\msun$ BH binaries are
too bright ($\sim 1$ mag) to account for the position of C1 on the
CMD. The disc flux may exceed the donor flux up to a factor of 10.
These models can therefore be excluded as possible candidates for
X-2. The tracks of 20$\msun$ BH binaries with donors of
$8\msun\simless M\simless 12\msun$ are compatible with the position of
C1 (with a relative flux contribution from the disc reaching $\sim
180$\% at maximum accretion rate). As for \textbf{case AB} mass
transfer, consistency with the photometry of C1, the observed value of
the orbital period and the age of field stars is never achieved for
donors of 8, 15, 20 and 25$\msun$. For the 10 and 12$\msun$ model, the
donor age is $\sim25$ and $18$Myr respectively when the orbital period
is $P_{orb}\sim 6.1$ d and the magnitude and colours are compatible
with C1. However, the binaries have spent only a tiny amount of time
($\sim 3000$ years for the 10 and 12$\msun$ donors) in RLOF contact
when all the parameters are within the observed range.

\section{Discussion}\label{discussion}

We demonstrated that, when using the CMD and the constraints from the
characteristic ages of the parent stellar cluster and the bubble
nebula, the binary evolution of X-2 in NGC 1313 is described
in terms of two possibilities for a \textbf{case AB} scenario:
\begin{itemize}
\item mass transfer of a MS donor of
$\sim12-15\msun$ onto a $\sim50-100\msun$ BH
\item mass transfer of a H-shell burning donor of 
 $\sim12-15\msun$ onto a $\sim20\msun$ BH
\end{itemize}

\textbf{Case B} mass transfer for a 50--100$\msun$ BH is finally ruled out as
the shift between the CMD tracks and the position of the optical
counterpart is too large. This result relies solely on the position
of the counterpart C1 on the CMD and assumes that all the binaries are
X-ray irradiated and that their emission is contaminated by the
accretion disc. Also \textbf{Case B} mass transfer for a 20$\msun$ BH
can be ruled out considering the too short-lived RLOF phase which 
appears in contradiction with the $\sim10^{6}$ yr age of the bubble 
nebula surrounding X-2 \citep{pak06}.
For \textbf{Case AB} mass transfer, the first contact phase during the MS is
sufficiently long to inject the required energy in the bubble nebula
and therefore makes the 20$\msun$ BH onto a 12-15 $\msun$ H-shell donor a 
possible candidate for X-2. However, a strongly beamed system would not
be consistent with the observations if the contribution to the nebular emission
from the photo-ionization by the ULX is important, as the nebula is essentially
isotropically illuminated \citep{pak06}.

Note that \textbf{case B} and \textbf{case AB} binary colours tend to
become redder as the evolution proceeds, as a consequence of surplus
mass ejection.  This is in contrast to what was reported by
\citet{mad08} who assumed that all the matter leaving the donor is
retained by the BH and contributes to the optical contamination.  When
the donor leaves the main sequence, the mass transfer rate increases
by up to 1-2 orders of magnitude. This means that in our model there
is a substantial mass loss from the binary that does not contribute to
the X-ray luminosity and to the reprocessing. This leads to redder
colours for the binary and to a stable mass transfer with a lack of a
common envelope phase (see \citealt{van05} and \citealt{lom05} for a
discussion).

If we use the tentative identification of the orbital period of
\citet{liu09} as \textit{further constraint}, binaries with a massive
50--100$\msun$ BH and a $\sim12$--$15\msun$ donor close to the TAMS
are compatible with the observations ($\sim6$ d), while all the
\textbf{case AB} binaries with a 20$\msun$ BH are excluded ($>8$ d).

In general the orbital modulation in the optical lightcurve is caused
by X-ray irradiation and ellipsoidal variations.  Simulating a
complete light curve is beyond the purpose of the present
investigation. As discussed in Section 4.1, we computed the amplitude
of the modulation induced by X-ray irradiation alone
($\sim0.05$--$0.15$ mag for $i \ga 45^0$). Ellipsoidal variations can
be estimated from \citet{boc79} and, for the typical mass ratios
considered here, have amplitudes $\sim0.1$--$0.2$ mag ($i \ga 45^0$).
So, the two effects are comparable. In both cases the reported values
refer to the amplitude of the modulation without considering the
accretion disc contribution.  If the latter is included, the amplitude
is reduced by a factor $\sim2$.  Ellipsoidal variations induce two
maxima and two minima per orbital cycle and therefore the observed
modulation of $\sim6$ days would correspond to an orbital period of 12
days.  However, depending on the parameters, the secondary minimum may
be largely suppressed and hence, within the photometric errors, the
observed modulation would correspond to the $\sim6$ days orbital period
adopted by \citet{liu09}. Clearly, more measurements are needed in
order to reach a definitive conclusion. If the orbital period were
$\sim12$ days, it would be compatible with a \textbf{case AB} binary
with a 20$\msun$ (or slightly larger) BH and isotropic irradiation
(8--20 d), while it would remain too small to be consistent with a
similar system in case beaming of the X-ray flux prevents irradiation
(15--23 d).

For 50--100$\msun$ BHs the expected amplitude of the orbital modulation is
larger in the $B$ band than in the $V$ or $R$ bands because, at longer
wavelengths, where the donor spectrum decays more rapidly than the
irradiated disc spectrum, the contamination from the disc is
comparatively stronger.  However, the difference in the amplitude
between the $B$ and $V$ band may not be sufficient to explain a
detection in the $B$ band and a simultaneous non-detection in the $V$
band. Further investigation is needed to assess this point. In the
present assumptions, the most favourable optical band where to search
for the orbital modulation appears to be the $U$ and $B$ bands, where the ratio
between donor and disc emission is maximum. It is interesting to note
also that, because of the disc contamination, the optical spectrum is
characterized by a rather flat continuum with $\propto \nu^{1.1}$,
clearly distinguishable from a Rayleigh-Jeans tail. All these
predictions may be easily tested with further photometric and
spectroscopic follow-ups of object C1.  Finally, if spectroscopic data
of the donor of NGC 1313 X-2 will support an identification of a MS
star, the possibility that NGC 1313 X-2 hosts a 50--100$\msun$ BH is strongly
favoured, independently of the period determination, whereas a giant
donor will immediately rule out the possibility of a BH this massive.
We note however that we did not perform a complete survey of the
parameter space evolving case AB systems with BH masses between 20 and
50$\msun$. In fact, depending on the actual value of the orbital period,
for values of the BH in this mass range, there may also be agreement
with observations. A systematic investigation of this type is postponed to
when a more robust assessment of the orbital period will be available.

\section*{Acknowledgments}
We would like to thank Lev Yungelson, Rudy Wijnands and Lex Kaper for
useful discussions. AP is supported by an NWO Veni fellowship. LZ
acknowledges financial support from INAF through grant PRIN-2007-26.

\newcommand{\nat}{Nat}
\newcommand{\mnras}{MNRAS}
\newcommand{\aj}{AJ}
\newcommand{\pasp}{PASP}
\newcommand{\aap}{A\&A}
\newcommand{\apjl}{ApJL}
\newcommand{\apss}{ApSS}
\newcommand{\apjs}{ApJS}
\newcommand{\aaps}{AAPS}
\newcommand{\apj}{ApJ}

\end{document}